\documentclass[twocolumn,showpacs,prl]{revtex4}
\usepackage{graphicx}

\begin{document}
\title{Theory of the tunneling resonances of the
bilayer electron systems in strong magnetic field}

\author{M. Abolfath$^{1,2}$}

\author{R. Khomeriki$^{1,3}$}

\author{K. Mullen$^1$}
\affiliation{$^{1}$University of Oklahoma, Department of Physics and Astronomy,
Norman OK 73019}
\affiliation{$^{2}$Department of Physics, University of Texas, Austin TX 78712}
\affiliation{$^{3}$Department of Physics, Tbilisi State University, Tbilisi
380028, Republic of Georgia}

\date{\today}

\begin{abstract}
We develop a theory for the anomalous interlayer
conductance peaks observed in  bilayer electron systems at $\nu=1$.
Our model shows the
that the size
of the peak at zero bias
decreases rapidly
with increasing in-plane magnetic field, but
its location is unchanged.
The I-V characteristic is linear at small voltages,
in agreement with experimental observations.
In addition we make quantitative predictions
for how the
inter-layer conductance peaks vary in position
with in-plane magnetic field at high voltages. Finally, we predict novel
bi-stable behavior at intermediate voltages.
\end{abstract}

\maketitle
The discovery of dual peaks in
the interlayer tunneling conductance in double layer quantum Hall systems
at total filling factor $\nu=1$ by
Spielman {\em et. al.} \cite{spielman1} has stimulated a
number of theoretical studies.\cite{WenZee,stern,schliemann,balents,fogler,mac}
Such interlayer tunneling measurements are a valuable tool
to study the dynamical
aspects of bilayer electron systems (BLES), since
the incident quasiparticles interact with the tunneling barrier, impurities,
interface roughness of the wells, as well as
the two-dimensional electron system
(2DES) in the wells.
The last of these gives rise to an inelastic scattering mechanism
where the tunneling quasiparticle
excites the collective modes of the BLES in strong magnetic field.
The dispersion relations of these modes,
and hence the inelastic scattering rate
are sensitive to the external in-plane field,
producing a resonant
peak in the conductance which varies in voltage with the applied
in-plane magnetic field.
The observation of this peak at non-zero voltage was been reported
by Spielman {\em et al.} \cite{spielman1}.
In addition, tunneling quasiparticles can interact with
topological defects in the order parameter such as
 merons (which carry the electrical
charge) causing phase decoherence and dissipation in the
tunneling current \cite{stern}, with
an inelastic scattering rate denoted by $\alpha_\perp$.
The height and the width of the interlayer current peak is limited by
such dissipative effects.

The least understood aspect
of the experiment
is the yet unexplained
peak in the conductance at zero voltage and its dependence
on the in-plane magnetic field.
The location of the
zero bias voltage peak is insensitive to the in-plane magnetic field,
but the height of this peak decreases rapidly with in-plane magnetic field.
Among the many theories of this
experiment\cite{WenZee,stern,balents,fogler,mac}
there are different and even
controversial interpretations for
these  observations.
For example in Refs. \cite{WenZee,fogler},
it has been argued that the conductance peak is
the remnant of the long Josephson effect, however,
in Ref. \cite{mac} the microscopic
calculation indicates this system can be described by the excitonic superfluids.

In this paper we propose an
approach based on a damped Landau-Lifshitz equation for the pseudospin
order parameter.  Our model differs from other approaches\cite{mac} in
the types of dissipation included and how interlayer current is calculated.
\cite{WenZee,stern,balents,fogler,mac}
These differences permit allow the model
to capture previously unexplained
features of the interlayer tunneling spectrum
in low and high bias voltages.
Our model reproduces
the experimentally observed
linear interlayer
current ($I_t \sim V$) at low voltages (fig. 1), along with the value of the
peak in the conductance at zero bias.
Taking the intrinsic damping mechanism into account,
we show the height of the conductance peak falls off as $1/Q^2$, where
$Q\equiv (B_\parallel/B_\perp) (d/\ell_0^2)$
is the in-plane magnetic field wave vector,
$\ell_0$ is the magnetic length, and $ B_\parallel$ ($B_\perp$)
is the component of the the field parallel (perpendicular) to the quantum
well.
At higher bias voltages the system enters a
non-linear regime with
a conductance peak whose position changes with in-plane magnetic field.
At still higher voltages, the current decays $I_t \sim 1/V^3$.
%
Furthermore, we argue the low voltage state is
different from the Josephson effect,
and therefore argue against possibility of the
Josephson effect at $V\rightarrow 0$.
Moreover, we predict the existence of a new
bistable state between a ``rotating" and ``locked" states for the order
parameter for small voltages in the presence of in-plane magnetic field,
could be realized, depending upon the initial conditions.


Our interpretation of the electrical current differs from other approaches.
We model the steady
state flow of quasiparticles, as an
{\em imperfect capacitor} with a non-linear charging energy, hence
the number of electrons and holes in different layers is fixed (but
not equal).
This  system of a parallel
resistor and capacitor (RC), connected to an external
electrochemical potential gives the steady state dissipative
interlayer current $I=\delta q(V)/\tau_z$.  Here $\delta q$ is the
restored charge in the capacitor (which is a function of external
potential $V$), and $\tau_z=RC$ is the relaxation time of the
circuit.  This analogy leads us to introduce a damping
coefficient $\alpha_z (\equiv 1/\tau_z)$ in the theory of the
interlayer tunneling effect.  This coefficient, absent in previous models,
is crucial to producing a zero bias peak.

\begin{figure}[t]
\begin{center}\leavevmode
\includegraphics[width=0.9\linewidth]{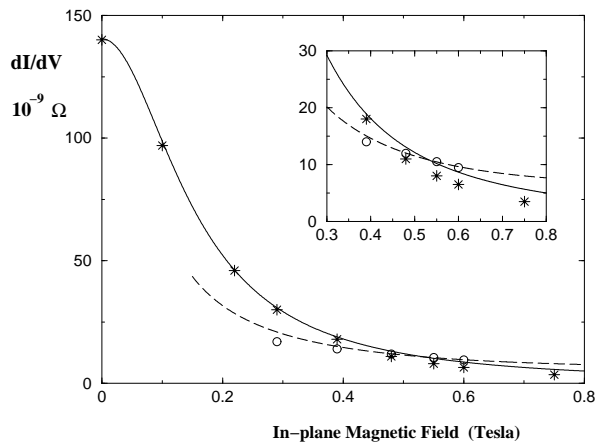}
\noindent
\caption{
The height of the zero voltage conductance peak (solid line)
and the conductance peak at non-zero voltage
(dashed line) versus the in-plane magnetic field,
obtained by Eqs. (\ref{12}) and (\ref{14}).
Stars and circles are the experimental data for
total electron density $N_T=6.0\times 10^{10}cm^{-2}$.
Inset: the magnified crossing region of the main panel.
The height of the zero-voltage peak falls off as $1/Q^2$;
the non-zero voltage peak varies as $1/Q$.  }
\end{center}\label{rrr}\vspace{-0.5cm}
\end{figure}

The low energy physics of the bilayer quantum Hall (pseudo) ferromagnets,
which is based upon the microscopic Hartree-Fock
model at odd integer filling fractions \cite{moon,yang},
is described by the following effective Hamiltonian:
\begin{eqnarray}
{\cal H}&=& - \frac{e V}{2} m_z + \frac{\rho_E}{2}\left\{\left(
\frac{\partial m_x}{\partial x}\right)^2
+\left(\frac{\partial m_y}{\partial x}\right)^2\right\}
+\beta m_z^2\nonumber \\ &&
-\Delta_{SAS}\Bigl\{m_x{\cos}(Qx)+m_y{\sin}(Qx)\Bigr\},
\label{1}
\end{eqnarray}
where $\hat m(x,t)$ is the order parameter unit vector ($m_z$ is the particle
density difference between two layers, and $m_x$, and $m_y$ are its
canonical conjugate variables), $\rho_E$ is the
in-plane (pseudo)spin stiffness, $\beta$ gives a hard axis anisotropy
due to the capacitance energy cost,
$\Delta_{SAS}$ is the tunneling amplitude,
$V$ is the external interlayer bias voltage, and $Q$ is defined above.

Without the external leads, $m_z=0$ is the lowest energy state of
an isolated BLES.
Connecting the layers to the external
leads bring this system out of equilibrium.
Similar to an imperfect capacitor, the quasi-particles can flow between
the layers, via a leakage current.
Before reaching steady state, the displacement current $dm_z/dt$,
(which passes through the capacitor even at $\Delta_{SAS}=0$), can be measured.
At steady state, the
charge density on the capacitor becomes fixed, even though there is still
a leakage current.
The capacitance charge is given by $<m_z (x, t)>$
($<>$ is the average over the temporal and the spatial fluctuations), which
is fixed and hence $<dm_z/dt>=0$.
However, there is still a steady state current due to
the interlayer quasiparticle tunneling channel.
(This result is
distinct from other approaches in which $<dm_z/dt>=0$ would imply there is
no tunneling current.)

The energy loss by dissipative quasiparticle tunneling
can be given phenomenologically by including
a damping coefficient $\alpha_z$, coupled to $m_z$.
\begin{equation}
I_t=\alpha_z e <m_z(x, t)>.
\label{7}
\end{equation}
The parameter $\alpha_z^{-1}$ which controls the resistance of the system
is equivalte to the RC relaxation rate, and
scales like $1/\sqrt{\beta\Delta_{SAS}}$.
To calculate $m_z$ we start from the {\em damped}
Landau-Lifshitz equations (see e.g. \cite{slichter}):
\begin{equation}
\vec R+\frac{\partial {\hat m}}{\partial t}=\bigl({\hat m}\times{\vec
H}_{eff}\bigr).
\label{3}
\end{equation}
${\vec H}_{\rm eff}=-2\left\{\frac{\partial{\cal H}}{\partial{\hat m}}
-\frac{\partial}{\partial x}\left[\frac{\partial{\cal H}}{\partial
\frac{\partial\hat m}{\partial x}}\right]\right\}$,
is the effective magnetic field, $\vec R$ is a Landau-Lifshitz damping term
(and $\hbar=1$).
Our phenomenological model for the damping in BLES
are characterized by two coefficients $\alpha_z$, and $\alpha_\perp$ that
define the damping vector $\vec R$:
\begin{eqnarray}
R_x&=&-\alpha_z m_z^2 m_x^0-\alpha_\perp m_y(m_y m_x^0-m_x m_y^0), \nonumber \\
R_y&=&-\alpha_z m_z^2 m_y^0+\alpha_\perp m_x(m_y m_x^0-m_x m_y^0), \nonumber \\&&
R_z=\alpha_z m_z(m_x m_x^0+m_y m_y^0).
\label{damp}
\end{eqnarray}
which is defined so that $\hat m \cdot \vec R=0$
and thus the length of $\hat m$
is conserved ($\hat{m}\cdot\partial_t\hat{m}=0$). The vector $\hat{m^0}$ is
the equilibrium value of the order parameter.
It is important to note the solutions of these Landau-Lifshitz equations
exhibit different behavior as the bias voltage is increased.
As $V \rightarrow 0$
(but $\Delta_{SAS} \ne 0$) the tunneling term is dominant and
the order parameter stays (almost) along the $x$-direction.
Without damping the order parameter can precess around this direction,
tracing out a cone centered on the $m_x$ axis.
The effect of the damping is to equilibrate
the order parameter along the $x$-direction
in a finite time, hence we assume $\hat{m}^0=(1,0,0)$ in Eq.(\ref{damp}).
Increasing the (small) bias voltage $V$  alters the equilibrium state.
Without damping, the lowest energy state can be determined by minimizing
Hamiltonian (\ref{1}).
The role of damping is to relax the excited states to some steady state
$\hat m^0$.
As $V$ increases, the
direction of $\hat m^0$ rotates towards the y-axis and tilts up slightly from
xy-plane.  The non-zero value of $m_y^0$ reflects a non-zero Josephson
current, but this current will vanish if $V=0$ due to the damping.  We refer
to this family of solutions as ``damping-locked states.''
%
Starting from $\hat{m}^0=(1,0,0)$ in Eq.(\ref{damp}), and
cranking up the bias voltages, when
$eV \approx \sqrt{8\beta\Delta_{SAS}}$,
the electrostatic energy  becomes comparable to the tunneling energy, and
the order parameter starts
to precess around a direction given by $\vec{H}_{\rm eff}$.
The system can no longer follow the tunneling term, and
the order parameter becomes ``unlocked'' due to the bias voltage.
The amplitude of these ``unlocked'' oscillatory solutions
decreases with increasing $V$, so that at very large bias voltages the order
parameter aligns with the z-axis as $V \rightarrow \infty$.
In the rest of the paper we detail the solutions of the
damped Landau-Lifshitz equations, and evaluate the interlayer current.
Starting from the uniform solution $m^0$, it is straightforward
to linearize the Landau-Lifshitz
equations, to find their inhomogeneous solutions,
using a starting point in our perturbative
expansion that is different in low and high $V$ limits
(due to different nature of solutions).

\begin{figure}[t]
\vspace{0.5cm}
\begin{center}\leavevmode
\includegraphics[width=0.8\linewidth]{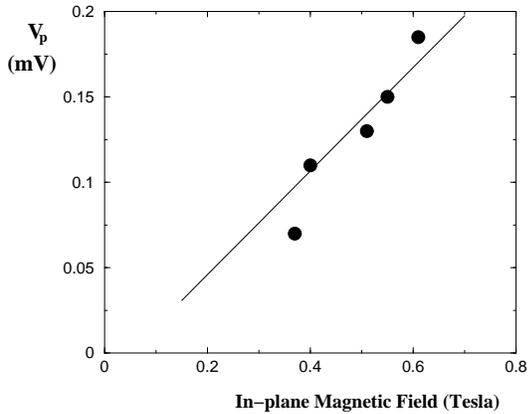}
\noindent
\caption{ Locations of non-zero voltage
conductance peaks versus in-plane magnetic field. The theoretical
curve (solid line) is derived from expression (\ref{14}), while the filled circles
are experimental data [7] for the same total electron density
$N_T=6.0\times 10^{10}cm^{-2}$.} \end{center}
\label{rrr}
\vspace{-0.5cm}
\end{figure}

\noindent {\it Large voltages:}
At high voltages ($eV \gg \sqrt{8\beta\Delta_{SAS}}$) the pseudospin
rotates around $z$-axis with a frequency $\omega \equiv eV/\hbar$.
It is then more convenient to work in a rotating frame
which can be introduced by the transformation
$n_\perp \equiv m_\perp \exp(i\omega t)$, $n_z \equiv m_z$, and
choosing $n^{(0)}=(1,0,0)$ which is equivalent to
$m^{(0)}=(\cos\omega t,\sin\omega t,0)$ in the rest frame.
Following this, the Hamiltonian (\ref{1}) in the rotating frame
can be transformed to
\cite{moon,yang}:
\begin{eqnarray}
&&{\cal H}=\frac{\rho_E}{2}\left\{\left(
\frac{\partial n_x}{\partial x}\right)^2
+\left(\frac{\partial n_y}{\partial x}\right)^2\right\}
+\beta (n_z)^2 \nonumber \\&&
-\Delta_{SAS}\Bigl\{n_x{\cos}(\omega t + Qx) + n_y{\sin}(\omega t + Qx)\Bigr\}.
\label{2}
\end{eqnarray}
Replacing $\hat n^0=(1, 0, 0)$ in Eq.(\ref{damp}),
the Landau-Lifshitz equations can be derived:
\begin{widetext}
\begin{eqnarray}
-{\alpha_\perp}n_y^2-\alpha_zn_z^2+\frac{\partial n_x}
{\partial t}=-4\beta n_yn_z-2\rho_E n_z
\frac{\partial^2 n_y}{\partial x^2}-2\Delta_{SAS}n_z{\sin}(\omega t+Qx) \nonumber \\
{\alpha_\perp}n_yn_x+\frac{\partial n_y}
{\partial t}=4\beta n_xn_z+2\rho_E n_z
\frac{\partial^2 n_x}{\partial x^2}+2\Delta_{SAS}n_z{\cos}(\omega t+Qx)
\label{5} \nonumber \\
\alpha_zn_zn_x+\frac{\partial n_z}
{\partial t}=2\rho_E\left\{n_x\frac{\partial^2 n_y}{\partial x^2}
-n_y\frac{\partial^2 n_x}{\partial x^2}\right\}+2\Delta_{SAS}\Bigl\{n_x{\sin}(\omega t+Qx)
-n_y{\cos}(\omega t+Qx)\Bigr\}.
\label{99}
\end{eqnarray}
\end{widetext}
The last equation in Eqs. (\ref{99}) can be interpreted as the
continuity equation for the interlayer current.
The external (but self-consistent) chemical potential contributes to the
current via the first term in the left side of this equation ($n_x \approx 1$).
The first term in the right hand side gives the current density due to phase
slips, $J=\rho_E \partial \varphi(x)/\partial x$, equivalent to
a dissipationless supercurrent density of the excitonic condensation.
In the presence of the small in-plane magnetic field (the commensurate state)
$J=\rho_E Q$.
Finally, the last term is analogous to the AC Josephson current.
The perturbative solution around $\hat n^0=(1, 0, 0)$ can be achieved by making the
harmonic expansion:
$\hat n=\vec A {\sin}(\omega t+Qx)+\vec B
{\cos}(\omega t+Qx)+\vec n^0 + \cdots$.
Substituting this into equations (\ref{5}), $A$ and $B$ can be determined
after linearizing the Landau-Lifshitz equation, and one can
derive the non-homogeneous leading terms in $\hat{n}(x,t)$.
Plugging this into Eq.(\ref{7}) (after replacing the coefficients $A$ and $B$ in
$\hat{n}(x,t)$), and making the space-time average,
we finally end up with an expression for the steady state tunneling DC current
\begin{eqnarray}
I_t=\frac{8e\beta \Delta_{SAS}^2\omega(\alpha_z+\alpha_\perp)}
{(8\beta\rho_E Q^2-\omega^2+\alpha_z\alpha_\perp)^2+\omega^2(\alpha_z+\alpha_\perp)^2}.
\label{14}
\end{eqnarray}
Solution (\ref{14}) qualitatively well describes the peaks at
$\omega \approx \sqrt{8\beta\rho_E} Q$,
corresponding to the resonance condition for the gapless acoustic mode.
The height and the
 width of the interlayer current are controlled by the damping.
In the absence of the damping $I_t (\propto \Delta^2_{SAS})$ has a singular peak
at $\sqrt{8\beta\rho_E} Q$.
Solution (\ref{14}) is parametrically unstable
for $\omega^2\geq 8\beta\rho_E Q^2$ (``tachionic" regime) as in the case of
long Josephson junctions, (see e.g. \cite{malomed}),
but it is stable for the large voltage limit
$\omega^2\gg8\beta\rho_E Q^2$ where the I-V characteristic follows the power law
$I_t\sim 1/V^3$.
Although it has been speculated by Fogler and Wilczek \cite{fogler} that the
interlayer current peaks
resemble the long AC Josephson effect \cite{eck} (where the location of the
peaks are shifted by $\alpha_\perp$, and $\alpha_z$),
here we argue the observation of these
peaks is the manifestation of the spontaneous phase coherence,
where the lowest energy state of the electrons is in a symmetric
linear combination of two layers which allow the electrons tunnel through the
energy barrier between two quantum wells without resistance
(if $\alpha_\perp = \alpha_z = 0$).
It is also possible to search for the excitonic superfluid modes
through out the interlayer tunneling measurement.
The staggered supercurrent density of the excitonic pairs in the
superfluid state
is given by $J_s = \rho_E Q_s$ for low tunneling energies.
The velocity of the collective modes alters by $J_s$:
$\omega_Q \rightarrow \omega_{Q+Q_s}$.
In the interlayer tunneling effect,
the incident electrons are scattered by these collective modes.
From the conservation of the energy and momentum, we have
$eV = \hbar \omega_{Q+Q_s}$, i.e., the superfluid current shifts the location
of the peaks.

Close to the order-disorder transition point, the experiment suggests
the possibility of the coexistence of the incompressible
state with the compressible state \cite{SternHalperin}.
In this circumstances, the current from one layer to other layer can transfer
through the phase coherent channel, as was described above, and also through the
quasiparticle channels which are in the ``uncorrelated'' state.
The contribution from the latter to the interlayer current in low voltages
is dominant, which gives rise to a linear I-V characteristic,
and therefore the possiblity of the DC Josephson effect is ruled out
(see below).
%

{\it Small Voltages:}
For low bias voltages, we begin with the Landau-Lifshitz equations,
Eqs.(\ref{3}), and (\ref{damp}) (in the rest frame).
Similar to Eqs.(\ref{99}) we can derive a set of equations in this limit.
The perturbative solution about $\hat m=(1,0,0)$ can be obtained
by a calculation similar to that above.
First we consider the simplest case of the zero
in-plane magnetic field ($Q=0$).
The uniform and static solution can be obtained easily.
Given these solutions, one can find the interlayer conductance 
\begin{equation}
G_t=\frac{2 e^2 \alpha_z \Delta_{SAS}}{4\Delta_{SAS}(2\beta+\Delta_{SAS})+
\alpha_z{\alpha_\perp}}. \label{9}
\end{equation}
A similar technique can be used to derive the analytical (non-uniform) solution
in the presence of in-plane magnetic field if $\rho_E Q^2\gg \Delta_{SAS}$.
The effect of the tunneling term (which is similar to the driving force)
is to create only
the harmonics with ``wave-number" $Q$ which itself will nonlinearly generate zero, $2Q$
and higher harmonics.
Because of the damping, the amplitudes of all other harmonics (but the first
harmonic) is expected to be zero in the zero temperature limit.
It is therefore natural to start with the following harmonic expansion
in the rest frame $m_x = 1 - m^+ m^- /2$, where
$m^+=A e^{iQx}+B e^{-iQx}+ m_0^+ + \cdots$, and $m^+ = m_z + i m_y$.
Assuming that the leading perturbative terms should be the first harmonics
of the driven wave-number
we can determine $A = -B = \Delta_{SAS} (4\beta - i \alpha_\perp)
/(8\beta\rho_E Q^2 + \alpha_z \alpha_\perp)$.
Substituting these into the Landau-Lifshitz equations, and linearizing
them in terms of $A$, and $B$, we can find the zeroth harmonic term
$m_0^+$, and then
the interlayer DC current $I_t=\alpha_z e <m_z(x,t)> = \alpha_z e m_{z0}$,
and the interlayer conductance can be obtained
\begin{eqnarray}
G_t =\frac{8 e^2 \beta \Delta_{SAS}^2\alpha_z}
{\alpha_z{\alpha_\perp} (8\beta\rho_E Q^2+\alpha_z\alpha_\perp)
+ (32\beta^2+2\alpha^2_\perp)\Delta^2_{SAS}}.
\label{12}
\end{eqnarray}
The height of this peak falls off like $1/Q^2$ (for high in-plane magnetic field),
but the location of its center does not vary with in-plane magnetic field,
consistent with \cite{spielman1} (see also Fig. 1).
This perturbative solution is valid for small $V$'s and large $Q$, and
it coincides with the
residual zero voltage peak in the presence of in-plane magnetic field.
%
%

We note in passing that these solutions are valid in their respective
limits, but that at intermediate voltages it may be possible to have more
than one solution to a nonlinear differential equation.  The basin of
attraction of the solutions will depend upon damping and other details of
the system.

{\em Numerical results:}
Our estimate shows the cross over between low and high bias voltages occurs
at $0.01 mV$.
To examine the accuracy of our model, the interlayer conductance
peaks have been drawn, by using two adjustable parameters.
In Figs. 1 and 2 the fit to the experimental
data \cite{spielman1} is obtained
by the following damping coefficients: $\alpha_\perp=0.25\alpha_z$,
$\alpha_\perp\alpha_z(\alpha_\perp+\alpha_z)^2=32\beta^2\Delta_{SAS}^2$
(for $\Delta_{SAS}=90~\mu K$, and
$\sqrt{8\beta\Delta_{SAS}}=70mK$, we find $\alpha_z=75mK$,
and $\alpha_\perp=18mK$).
In Fig. 1, the height of the central residual (solid line)
and the split off peaks (dashed line) vs. $B_\parallel$ have been
derived by Eq. (\ref{14}), and (\ref{12}). 
In Fig. 2 we present the locations of the split off peaks derived
from Eq. (\ref{14}).

We presented a physical picture
based on a driven-damped easy-plane pseudospin
ferromagnet model for the
experimental observation of the interlayer conductance peak in
a bi-layer electron system at $\nu=1$ \cite{spielman1}.
The first theoretical prediction for low bias voltage
conductance peak vs. in-plane magnetic field has been made.
It has been shown, at high voltages, due to the non-linear
behavior of the capacitance energy, the inter-layer current shifts by
in-plane magnetic field.

We gratefully acknowledge helpful interactions with Allan MacDonald,
Steve Girvin, and Elena Bascones.
This work is supported by NSF MRSEC DMR-0080054 (MA),
NSF-NATO DGE-0075191 (RK) and NSF EPS-9720651 (KM).

\end{document}